\documentclass[aps,showpacs,showkeys,preprintnumbers,amsmath,epsf,amssymb,epsfig,nofootinbib]{revtex4}
\input epsf.sty
\usepackage{bm}
\usepackage{graphicx}
\draft
\begin{document}

\title{Phantom cosmology with a decaying cosmological function $\Lambda(t)$ induced from five-dimensional (5D) geometrical 
vacuum}
\author{ $^{1}$ Jos\'e Edgar Madriz Aguilar \footnote{
E-mail address:jemadriz@fisica.ufpb.br }, $^{2,3}$
Mauricio Bellini \footnote{E-mail address: mbellini@mdp.edu.ar} and $^{1}$ Marco A. S. Cruz \footnote{
E-mail address:mcruz@fisica.ufpb.br }}

\address{$^{1}$ Departamento de F\'{\i}sica,\\ Universidade
Federal da Para\'{\i}ba. C. P. 5008, CEP 58059-970 Jo\~{a}o Pessoa, Para\'{\i}ba-Brazil.\\
$^{2}$Departamento de
F\'{\i}sica, Facultad de Ciencias Exactas y Naturales, Universidad
Nacional de Mar del Plata, Funes 3350, (7600) Mar del Plata,
Argentina.\\
$^3$ Consejo Nacional de Investigaciones
Cient\'{\i}ficas y T\'ecnicas (CONICET).
}

\begin{abstract}
Introducing a variable cosmological function $\Lambda (t)$ in a
geometrical manner from a 5D Riemann-flat metric, we investigate
the possibility of having a geometrical criterion to choose a
suitable cosmological function $\Lambda (t)$ for every 4D
dynamical hypersurface capable of generate phantom cosmologies.
\end{abstract}


 \maketitle
\section{Introduction}

The observational cosmological data indicate that the present values for dark energy
and matter components in terms of a critical energy density are approximately $\Omega_{\Lambda}
\simeq 0.7$ and $\Omega_{M} \simeq 0.3$ \cite{pdg}. This observational evidence is compelling
for a spatially flat and low matter density universe which nowadays is in a period of accelerated
expansion. An immediate consequence of this observational scenario is that the cosmological fluid
seems to be dominated by some sort of dark energy. This kind of fantastic energy that remains unclustered at all scales and has 
negative pressure plays a predominant role in the present epoch. Some of the candidates postulated for modeling dark energy are  
the cosmological constant $\Lambda_0$ having an
equation of state $\omega_{eff}=P_{eff}/\rho_{eff} =-1$ ($P_{eff}$
and $\rho_{eff}$ are respectively the effective pressure and
energy density of the universe), and the so called quintessence models which typically leads to an equation of state parameter 
ranging in the interval $-1 <\omega_{eff} < -1/3$. However in the case of the cosmological constant the $\Lambda$CDM model failed 
in explaining why the inferred value of $\Lambda _0$ is so tiny (120 orders of magnitude lower) compared to the typical vacuum 
energy values inferred by particle physics (the coincidence problem) \cite{Cap}. In spite of being successful in fitting the 
observational data in the majority of quintessential models the coincidence problem even prevails.  Another interesting class of 
models that also fits the observational data are the called phantom
cosmologies \cite{PC}. On this phantom scenarios the equation of state parameter satisfies
$\omega_{eff} < -1$. The phantom energy component predicted in some of this models
is compatible with most classical tests of cosmology based on the
current data including the type SNeIA data as well as the cosmic
microwave background anisotropy and mass power spectrum.\\

On the other hand, during the last years theories in more than
four dimensions have become one of the fundamental cornerstones
of modern physics. The most appealing are string theories \cite{Dim1}, braneworld
inspired models \cite{RS12}, the class of Kaluza-Klein theories and among the last ones the Induced Matter theory (IMT) is counted
\cite{Wess1}. The IMT is based on the assumption that ordinary
matter and physical fields that we can observe in our 4D universe can be geometrically induced from a 5D space-time in vacuum due 
to the existence of a noncompact extra dimension \cite{im}. On this framework, inflationary
models induced from a 5D vacuum state where the expansion of the
universe is driven by a single scalar (inflaton) field have been
subject of great activity during the last years \cite{nuestros}.\\

In this letter our interest is to investigate the possibility of
obtaining phantom cosmological scenarios from a 5D vaccum state.
In order to achieve it we use a dynamical foliation on the
space-like fifth coordinate, here considered as noncompact. The
letter is organized as follows: in Sect. II we introduce the
cosmological function in a geometrical manner. In Sect. III we
study three examples; in the first one we study an expansion with
a constant cosmological function In the second we study a decaying
cosmological
function extracting a quintessential scenario while in the third example we obtain a phantom scenario from a decaying 
cosmological function Finally, in Sect. IV we give some final remarks.\\

\section{Introducing a cosmological function $\Lambda (t)$ in a geometrical manner}

In order to introduce a variable cosmological function $\Lambda
(t)$ in a geometrical manner we consider the recently introduced
5D line element \cite{MET}
\begin{equation}\label{a1}
dS^{2}=\psi^{2}\,\frac{\Lambda(t)}{3}\,dt^{2}-\psi ^{2}
\,e^{2\int \sqrt{\Lambda(t)/3}\,dt}dr^{2}-d\psi ^{2},
\end{equation}
where $dr^{2}=\delta _{ij}dx^{i}dx^{j}$ is the 3D Euclidean
metric, $t$ is the cosmic time and $\psi$ is the space-like extra
dimension. Choosing a natural unit system the cosmological
function $\Lambda (t)$ has units of $[length]^{-2}$. After some
straightforward calculations it can be easily shown that the
metric $g_{AB}$ in (\ref{a1}) is Riemann-flat and consequently
Ricci-flat,
making it a suitable metric for describing a 5D vacuum. Note that we are working within the context of the induced matter theory 
and the Riemann flatness of the metric (\ref{a1}) do not modify in any way the mechanism here proposed. In other words this 
mechanism is valid in general for every Ricci-flat metric containing $\Lambda (t)$ and the metric (\ref{a1}) must be taken as a 
particular solution of $R_{AB}=0$. \\

Now let us to assume that the 5D space-time can be
dynamically foliated with generic dynamical hypersurfaces
$\Sigma:\psi=f(t)$. Thus the line element (\ref{a1}) on $\Sigma$ becomes
\begin{equation}\label{a2}
dS_{\Sigma}^{2}=\left[f^{2}(t)\frac{\Lambda(t)}{3}-
\dot{f}^{2}(t)\right]dt^{2}-f^{2}(t)\,e^{2\int \sqrt{\Lambda(t)/3}\,dt}dr^{2},
\end{equation}
with the dot denoting derivative with respect to the time $t$.
Adopting the continuity conditions introduced by J. Ponce de Leon in \cite{JPLeon} to get a 4D FRW-metric on $\Sigma$, we obtain
\begin{eqnarray}\label{a3}
f^{2}(t)\frac{\Lambda(t)}{3}-\dot{f}^{2}(t)&=&1\\
\label{a4}
f^{2}(t)\,e^{2\int \sqrt{\Lambda(t)/3}\,dt}&=&a^{2}(t)\, ,
\end{eqnarray}
being $a(t)$ an effective scale factor describing
the 3D-spatial expansion on $\Sigma$. In general, solutions of equation (\ref{a3}) determine a family of hypersurfaces
$\lbrace\Sigma:\psi =f(t)\rbrace$ where the line element (\ref{a2})
is valid whereas equation (\ref{a4}) specifies an effective scale factor
$a(t)$ for every element of the family.\\

On a generic hypersurface $\Sigma:\psi=f(t)$ the
induced 4D Einstein equations read
\begin{eqnarray}
3\left(\frac{\dot{a}}{a}\right)^{2}&=&k\,\rho _{IM}+\Lambda, \label{v1} \\
2\frac{\ddot{a}}{a}+\left(\frac{\dot{a}}{a}\right)^{2}&=&-(k\,P_{IM}-\Lambda),\label{v2}
\end{eqnarray}
where $k=8\pi G$, being $G$ the Newton's constant and the induced
matter density $\rho _{IM}$ and pressure $P_{IM}$ are given by
\begin{eqnarray}\label{v3}
k\rho _{IM}&=&3\left(\frac{\dot{f}}{f}\right)^{2}
+2\frac{\dot{f}}{f}\sqrt{3\Lambda},\\
\label{v4}
kP_{IM}&=&-2\frac{\ddot{f}}{f}-2\sqrt{3\Lambda}\frac{\dot{f}}{f}-
\frac{\dot{\Lambda}}{\sqrt{3\Lambda}}-\left(\frac{\dot{f}}{f} \right)^{2}.
\end{eqnarray}
The system (\ref{v1})-(\ref{v2}) can be interpreted as the
dynamical equations of a Friedmann-Robertson-Walker (FRW)
cosmology with a 4D energy momentum tensor given by
$T_{\mu\nu}=(\rho _{IM}
+P_{IM})\,u_{\mu}u_{\nu}-P_{IM}\,g_{\mu\nu}+k^{-1}\Lambda(t)
g_{\mu\nu}$, being $u_{\mu}$ the 4-velocities associated to the
4D-comoving observers on every dynamical hypersurface $\Sigma:\psi
= f(t)$. In other words we have induced from 5D-vacuum an
energy-momentum tensor that can be interpreted as an
energy-momentum tensor describing a perfect fluid under the
presence of a cosmological function $\Lambda(t)$. This
energy-momentum tensor can be written in terms of an effective
energy density $\rho _{eff}$ and an effective pressure $P_{eff}$
as $T_{\mu\nu}^{(eff)}=(\rho
_{eff}+P_{eff})\,u_{\mu}u_{\nu}-P_{eff}\,g_{\mu\nu}$, being $\rho
_{eff}$ and $P_{eff}$ defined by
\begin{equation}\label{v5}
\rho _{eff}=\rho _{IM}+k^{-1}\Lambda(t),\qquad P_{eff}=
P_{IM}-k^{-1}\Lambda(t).
\end{equation}
Using (\ref{v3}), (\ref{v4}) and (\ref{v5}) we can establish an
effective equation of state (EOS) of the form $P_{eff}=\omega
_{eff}\,\rho _{eff}$, where the EOS parameter $\omega _{eff}$ is
determined by
\begin{equation}\label{v6}
\omega _{eff} (f)=-\left[1+\frac{2(\ddot{f}/f)-2(\dot{f}/f)^{2}
+(\dot{\Lambda}/\sqrt{3\Lambda})}{3(\dot{f}/f)^{2}
+2\sqrt{3\Lambda}(\dot{f}/f)+ \Lambda}\right].
\end{equation}
Employing equation (\ref{a4}) the induced deceleration parameter
$q=-(\ddot{a}a/\dot{a}^2)$ on the brane $\Sigma: \psi =f(t)$
acquires the form
\begin{equation}\label{pv7}
q_{eff}(f)=-\frac{f\left[\ddot{f}+2\sqrt{\Lambda/3}\,\dot{f}
+(1/2)(\dot{\Lambda}/\sqrt{3\Lambda})f+(\Lambda/3)f\right]}{\left[ \dot{f}+ \sqrt{ \Lambda / 3}\,f\right]^2}.
\end{equation}
By simple inspection of expression (\ref{a3}) it can
be easily seen  that when we assume {\it a priori}
constant foliations and consider the case of a cosmological constant  $\Lambda
=\Lambda _{0}$ the corresponding constant foliations are determined by $f=f_{0}=\pm \sqrt{3/\Lambda _{0}}$. Then it
follows from (\ref{v6}) and (\ref{pv7}) that the effective
EOS parameter and the deceleration parameter become $\omega _{eff}=-1$ and $q=-1$
respectively, describing in this way a perfect vacuum EOS. Moreover from expression (\ref{a4}) the corresponding scale factor
is $a(t)=f_{0}\exp[\sqrt{\Lambda _{0}/3}\,t]$
which indicates that under that conditions we can recover a de-Sitter expansion.\\

\section{Examples}

Another
immediate implication that arises from equation (\ref{a3}) is that if we regard ({\it a priori}) a constant cosmological function 
$\Lambda
=\Lambda _{0}$, in principle it is possible to have a dynamical foliation $f=f(t)$ that satisfies (\ref{a3}). These
cases are the subject of the following examples.

\subsection{Inducing a ''constant'' cosmological function}

The simplest case that we can consider in our analysis is when the
cosmological function $\Lambda(t)=\Lambda _{0}$ is constant. In
such a case the continuity conditions (\ref{a3}) and (\ref{a4})
yield
\begin{eqnarray}\label{a5}
f^{2}(t)\frac{\Lambda _{0}}{3}-\dot{f}^{2}(t)&=&1 \, ,\\
\label{a6}
f(t)e^{\sqrt{\Lambda _{0}/3}\,t}&=&a(t)\,.
\end{eqnarray}
Solving (\ref{a5}) we obtain
\begin{equation}
\label{a7}
f_{1}(t)=\pm\sqrt{\frac{3}{\Lambda _0}},\quad f_{2}(t)
=\frac{1}{6}\sqrt{\frac{3}{\Lambda _0}}\left[9\,e^{-\sqrt{\Lambda _0/3}\,(t-t_{0})}+e^{\sqrt{\Lambda _0/3}\,(t-t_{0})}\right],
\end{equation}
\begin{equation}
\label{a8}
\quad f_{3}(t)=\frac{1}{6}\sqrt{\frac{3}{\Lambda _0}}
\left[9\,e^{\sqrt{\Lambda _0/3}\,(t-t_{0})}+e^{-\sqrt{\Lambda _0/3}\,(t-t_{0})}\right],
\end{equation}
being $t_{0}$ an integration constant that could be
interpreted as some initial time. Clearly $f_{1}$ corresponds to
a couple of constant foliations while $f_{2}$ and $f_{3}$ give dynamical
ones. The Friedmann equations (\ref{v1}) and (\ref{v2}) read
\begin{eqnarray}\label{a9}
3H^{2}&=&k\,\rho _{IM_{(0)}}+\Lambda _{0}\\
\label{a10} 2\frac{\ddot{a}}{a}+\left(\frac{\dot{a}}{a}\right)^{2}
&=& -(k \,P_{IM _{(0)}}-\Lambda _0),
\end{eqnarray}
where $\rho _{IM _{(0)}}=3(\dot{f}/f)^{2} + 2(\dot{f}/f)
\sqrt{3\Lambda _0}$ and $P_{IM _{(0)}}=-2(\ddot{f}/f)-
2\sqrt{3\Lambda _0} - (\dot{f}/f)^{2}$ are the induced energy density and pressure
in the presence of $\Lambda _0$. As we mentioned in the previous section in the case of the hypersurfaces $f_{1}=
\pm\sqrt{3/\Lambda _0}$ corresponding EOS parameter
and the deceleration parameter are just $\omega _{eff}(f_1)=-1$ and $q_{eff}(f_{1})=-1$.
However in the cases of $f_{2}$ and $f_{3}$ they become
\begin{eqnarray}\label{a11}
&&\omega _{eff}(f_{2})=-\left[1+6e^{-2\sqrt{\Lambda _0 /3}(t-t_{0})}
\right],\qquad\omega _{eff}(f_{3})=-\left[1
+\frac{2}{27}\,e^{-2\sqrt{\Lambda _0 /3}(t-t_{0})}\right],\\
\label{a12}
&& q_{eff}(f_{2})=-\left[1+9e^{-2\sqrt{\Lambda _{0}/3}(t-t_0)}
\right],\qquad q_{eff}(f_3)=-\left[1+\frac{1}{9}\,e^{-2\sqrt{\Lambda _0 /3}(t-t_0)}\right].
\end{eqnarray}
From these expressions we can see immediately that
when $t-t_{0} \gg 1$, all of them tend asymptotically
to a de-Sitter cosmology.
Or in geometrical terms we can say that in the present
case the asymptotic limit of the dynamical foliations are the constant ones.

\subsection{Inducing a variable cosmological function}

The results discussed in the previous sections have been obtained
interpreting the continuity condition (\ref{a3}) as an equation
whose solutions determine the dynamical foliations on which the
line element (\ref{a2}) holds. Something remarkable to find
solutions of (\ref{a3}) is that we must know {\it a priori} an
algebraic expression for the cosmological function $\Lambda (t)$
or having an extra criterion to provide it. However, we can
interpret equation (\ref{a3}) from another point of view. We can
regard expression (\ref{a3}) as a criterion of choosing or
inducing cosmological functions when we give a suitable family of dynamical hypersurfaces that satisfies (\ref{a3}). In 
mathematical
terms, we mean that we can express equation (\ref{a3}) as
\begin{equation}\label{Ma1}
\Lambda (t)=\frac{3}{f^2(t)}\left[1+\dot{f}^{2}(t)\right].
\end{equation}
From this point of view for a given hypersurface
$\Sigma:\psi =f(t)$ or dynamical foliation we have an induced
$\Lambda (t)$ specified by the expression (\ref{Ma1}). The freedom that
we have now for choosing a suitable dynamical foliation can be reduced if we inspire our election in the physical situation or the
issue addressed with this formalism. In particular we are interested in cosmological applications and in particular in study
the epochs of the universe where the presence of a
dynamical  cosmological is preponderant.\\

\begin{figure}
\includegraphics[width=10cm]{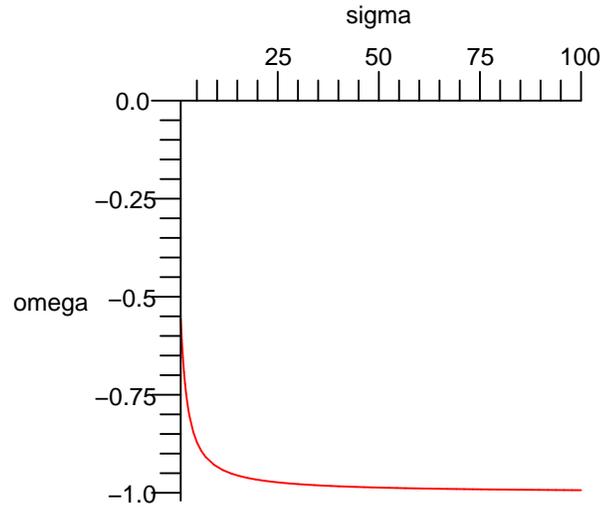}
\caption{\label{fig1} The figure shows the behavior of the effective EOS parameter $\omega _{eff}$ with the parameter $\sigma$.}
\end{figure}

\begin{figure}
\includegraphics[width=9cm]{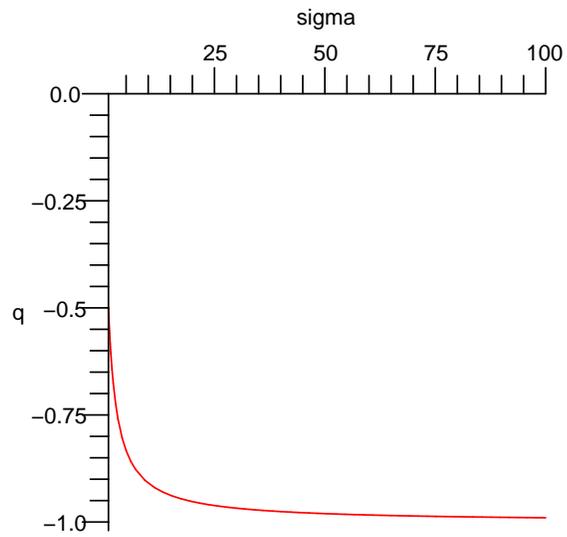}
\caption{\label{fig2} The figure shows the behavior of the effective deceleration parameter $q _{eff}$ with the parameter 
$\sigma$.}
\end{figure}

Thus, in order to illustrate the formalism let us
to consider the
particular cases when we have  $f=H_{0}^{-1}$ and  $f=\alpha^{-1}t$, being $\alpha$ a nonzero constant parameter and $H_{0}$ the 
constant Hubble parameter. Thus the induced cosmological functions
$\Lambda (t)$ according to (\ref{Ma1}) are respectively
\begin{equation}\label{b4}
\Lambda _{c} =3H_{0}^{2},\qquad\qquad \Lambda (t)=\frac{3(\alpha^{2}+1)}{t^{2}}.
\end{equation}
Using equation (\ref{a4}) it follows that the effective
scale factors in both cases are given by
\begin{equation}\label{b5}
a(t)=H_{0}^{-1}e^{H_{0}(t-t_{i})},\qquad a(t)
=\frac{1}{\alpha}\,t_{i}^{-(1+\alpha^2)^{1/2}}t^{1+\sqrt{1+\alpha^2}},
\end{equation}
where $t_{i}$ is some initial time determined as an initial
condition. As it is natural of being expected the first scale factor corresponds to a de Sitter expansion while the second
one correspond a power-law type expansion. Clearly the case of a constant Hubble parameter corresponds to a constant foliation, so
that as we have shown before in that case both the effective EOS parameter and the effective deceleration parameter become -1
describing a de-Sitter expansion. However the case of power law expanding universe is more interesting. In such
a case the $\omega _{eff}$ and $q_{eff}$ read
\begin{eqnarray}\label{b6}
\omega _{eff}&=& -\left[1-\frac{2}{3}\left(\frac{\sigma +1}{\sigma ^2
+ \sigma +1}\right)\right],\\
q_{eff}&=& -\frac{\sigma }{\sigma +1}
\end{eqnarray}
being $\sigma=\sqrt{\alpha^{2}+1}$. The behavior of $\omega
_{eff}$ and $q_{eff}$ is plotted in the figures [\ref{fig1}] and
[\ref{fig2}]. Notice that both, $\omega_{eff}$ and $q_{eff}$
remain above $-1$ for $\alpha \geq 0$ but always are negative. More specifically, for
$\sigma \geq 1$ we obtain: $-1 \leq \omega_{eff} <-5/9 $ and
$-1 \leq q_{eff} < -1/2$, which means that we are mimicking a period of quintessential expansion. On the other hand, the fact that  
the both parameters $\omega
_{eff}$ and $q_{eff}$ remain always negative is interpreted here as the metric
(\ref{a1}) in the present mechanism can only describe phases in
the evolution of the universe which are dominated by vacuum
energy, or equivalently by a cosmological function $\Lambda (t)$. In more general terms we can say that under the consideration 
of foliations like $f=\alpha^{-1}t$ the metric in (\ref{a1}) is able of describing only phases of accelerated expansion of the 
universe.

\subsection{Extracting a phantom cosmological scenario}

In this section our interest is to extract some phantom cosmological scenarios from a geometrically induced $\Lambda (t)$. This 
cosmological function $\Lambda$ must be a decreasing function of time in order to explain its tiny value today \cite{carneiro}. 
Phantom cosmology is an alternative to explain the present accelerated expansion of the universe characterized by an EOS parameter 
$\omega _{ph}<-1$. As it was shown in \cite{Chatterjee} within the context of 5D cosmological theories solutions for the scale 
factor involving hyperbolic functions can describe an accelerated expansion and in particular the present one. On the other hand, 
we must note that from (\ref{a4}) we can derive the expression $H=(\dot{f}/f)+\sqrt{\Lambda/3}$, so it results clear given that 
$\Lambda$ is written in terms of the dynamical foliation $f(t)$ that if we want to obtain a scale factor $a(t)$ in terms of 
hyperbolic functions like in \cite{Chatterjee} the dynamical foliation $f$ must be also given in terms of hyperbolic functions. 
Thus in order to show the possibility to extract phantom cosmological scenarios by using the formalism here discussed, let us try 
the ansatz for the dynamical foliation $f(t)$ given by
\begin{equation}\label{ext1}
f(t)=\beta \cosh[\bar{\alpha}(t-t_{\star})],
\end{equation}
where $t_{\star}$ is the time when the period of accelerated expansion to be mimicked begins and $\beta$ and $\bar{\alpha}$ being  
not null constant parameters having $\beta$ units of {\it length} and $\bar{\alpha}$ units of {\it (length) $^{-1}$} (Note that 
here we are using natural units). Thus it follows from the equation (\ref{Ma1}) that the induced cosmological function has the 
form
\begin{equation}\label{ext2}
\Lambda (t)=3\left[\left(\bar{\alpha}^{2}-\frac{1}{\beta^2}\right)\tanh^{2}[\bar{\alpha}(t-t_{\star})]+\frac{1}{\beta ^2}\right],
\end{equation}
which for values of $\beta$ and $\bar{\alpha}$ satisfying $(\bar{\alpha}\beta)^{2} < 1$ is a decreasing function of time. The 
induced Hubble parameter according to (\ref{a4}) is in this case
\begin{equation}\label{ext3}
H(t)=\bar{\alpha}\,\tanh[\bar{\alpha}(t-t_{\star})]+\left[\frac{1}{\beta 
^2}sech^{2}[\bar{\alpha}(t-t_{\star})]+\bar{\alpha}^{2}\tanh^{2}[\bar{\alpha}(t-t_{\star})]\right]^{1/2}.
\end{equation}
Thus inserting (\ref{ext1}) and (\ref{ext2}) in (\ref{v6}) and (\ref{pv7}) the effective EOS parameter and the deceleration 
parameter result respectively
\begin{eqnarray}\label{ext4}
\omega _{eff}(t)&=& 
-\frac{1}{3}\frac{2\bar{\alpha}^{2}\beta^{2}\sqrt{1+\bar{\alpha}^{2}\beta^{2}\sinh^{2}[\bar{\alpha}(t-t_{\star})]}\,\left(3\cosh^{
2}[\bar{\alpha}(t-t_{\star})]-2\right)+g_{1}(t)}{g_{2}(t)\,\sqrt{1+\bar{\alpha}^{2}\beta^{2}\sinh^{2}[\bar{\alpha}(t-t_{\star})]}}
,\\
\label{ext5}
q_{eff}(t)&=&-\frac{\bar{\alpha}^{2}\beta^{2}\left(2\cosh^{2}[\bar{\alpha}(t-t_{\star})]-1\right)+1}{\left(\,\bar{\alpha}\beta 
\sinh[\bar{\alpha}(t-t_{\star})]+\sqrt{1+\bar{\alpha}^{2}\beta^{2}\sinh^{2}[\bar{\alpha}(t-t_{\star})]}\,\right)\,\sqrt{1+\bar{ 
\alpha}^{2}\beta^{2} \sinh^{2}[\bar{\alpha}(t-t_{\star})]}},
\end{eqnarray}
being
\begin{eqnarray}\label{ext6}
g_{1}(t)&=& 4\bar{\alpha}\beta \sinh [\bar{\alpha}(t-t_{\star})]+6\bar{\alpha}^{3}\beta 
^{3}\sinh[\bar{\alpha}(t-t_{\star})]\cosh^{2}[\bar{\alpha}(t-t_{\star})]-4\bar{\alpha}^{3}\beta^{3}\sinh[\bar{\alpha}(t-t_{\star})
],\\
\label{ext7}
g_{2}(t)&=& 2\bar{\alpha}^{2}\beta 
^{2}\cosh^{2}[\bar{\alpha}(t-t_{\star})]-2\bar{\alpha}^{2}\beta^{2}+2\bar{\alpha}\beta\sinh[\bar{\alpha}(t-t_{\star})] 
\sqrt{1+\bar{\alpha}^{2}\beta ^{2}\sinh^{2}[\bar{\alpha}(t-t_{\star})]}+1.
\end{eqnarray}
Now in order to show that there exist suitable values of $\bar{\alpha}$ and $\beta$ leading to a phantom scenario mimicked 
geometrically by the use of dynamical foliations of the 5D space-time, let us to consider the values $\bar{\alpha}=1/2$ and $\beta  
=3/2$. With these values the temporal behavior of $\omega _{eff}$ and $q_{eff}$ is plotted in the figures [\ref{fig3}] and 
[\ref{fig4}]. In the figures we can see that we mimicked an EOS passing from a period of phantom cosmology which tends 
asymptotically to a de-Sitter stage.

\section{Final Comments}

In this letter we have derived phantom cosmologies from a 5D Riemann-flat space-time with the metric ansatz (\ref{a1}) and where 
the fifth coordinate has been regarded as noncompact. Our formalism is based on the assumption that the 5D space-time can be 
dynamically foliated by a family of 4D hypersurfaces $\Sigma: \psi =f(t)$. In order to induce a FRW line element on every 4D 
hypersurface  we have adopted the continuity conditions of Ponce de Leon (\ref{a3}) and (\ref{a4}). On that framework we have 
shown that the energy momentum tensor geometrically induced on every dynamical hypersurface  can perfectly describe a perfect 
fluid under the presence of a variable cosmological function $\Lambda (t)$. In this sense the present formalism can be considered 
as a way of introducing a cosmological function $\Lambda (t)$ geometrically from pure 5D vacuum. By reinterpreting the condition 
(\ref{a3}) we have established a geometrical criterion for assigning cosmological functions $\Lambda (t)$ which depends only of 
the dynamical foliation adopted.\\

When we regard the foliation $\Sigma:\psi =f(t)$, the induced 4D dynamics is governed by the system (\ref{v1})- (\ref{v2}). When 
the condition (\ref{a5}) is adopted it can be easily seen that this induced dynamics describes a 4D universe with an effective 4D 
FRW metric $ds^2=dt^2-a^2 dr^2$. Under this conditions a quintessential expanding period in the evolution of the universe can be 
mimicked. In this case we obtain a 4D effective EOS parameter $\omega_{eff}=P_{eff}/\rho_{eff}$ and a deceleration parameter 
$q_{eff}$ range between $-1\leq \omega _{eff}<-5/9$ and $-1\leq q_{eff}<-1/2$, respectively, when $\sigma \geq 1$. One interesting 
feature of both $\omega _{eff}$ and $q_{eff}$ is that they remain negative and always above $-1$ for $\sigma \geq 1$. This 
characteristic suggests that under that conditions the metric ansatz (\ref{a1}) can describe only periods of accelerated expansion 
of the universe dominated by a dark energy here mimicked by a geometrically induced cosmological function $\Lambda (t)$.\\

\begin{figure}
\includegraphics[width=9cm]{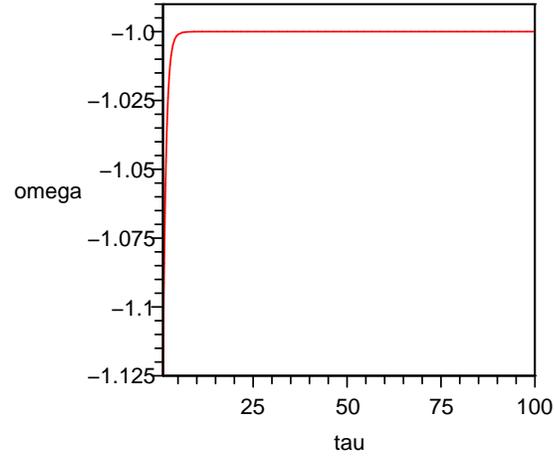}
\caption{\label{fig3} The figure shows the behavior of the effective EOS parameter $\omega _{eff}$ with $\tau =t-t_{\star}$ when 
the values $\bar{\alpha} =1/2$ and $\beta =3/2$ are considered.}
\end{figure}

\begin{figure}
\includegraphics[width=8cm]{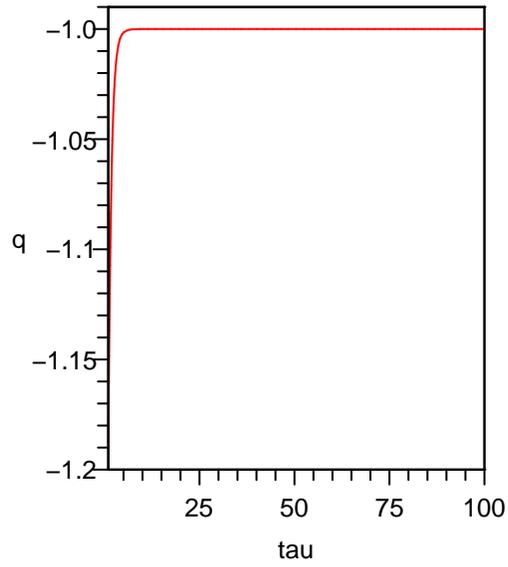}
\caption{\label{fig4} The figure shows the behavior of the effective deceleration parameter $q _{eff}$ with $\tau =t-t_{\star}$ 
when the values $\bar{\alpha} =1/2$ and $\beta =3/2$ are considered.}
\end{figure}

Finally, phantom cosmological scenarios can be also obtained by choosing a suitable dynamical foliation. More precisely we can 
obtain phantom scenarios from a geometrically induced cosmological function $\Lambda (t)$. In this case regarding for instance 
the particular foliation $f(t)=\beta \cosh[\bar{\alpha}(t-t_{\star})]$ we can induce for $(\bar{\alpha}\beta)^{2}<1$ a decaying 
cosmological function $\Lambda (t)$ that can lead to a phantom
scenario. Other possible explanation to solve the called
cosmological constant problem, but in 4D, was developed in
\cite{MK}.
In particular for $\bar{\alpha}=1/2$ and $\beta =3/2$ we 
can achieve a phantom scenario which tends asymptotically to a de-Sitter expansion. The behavior of $\omega _{eff}$ and $q_{eff}$ 
for this case with respect to $\tau =t-t_{\star}$ is plotted in
figures [\ref{fig3}] and [\ref{fig4}] respectively.

\begin{acknowledgments}
 M. A. C. S. and J. E. M. A. acknowledge CNPq-CLAF and UFPB (Brazil)
and MB acknowledges CONICET and UNMdP (Argentina) for financial support.
\end{acknowledgments}

\newpage

\newpage

\end{document}